\documentclass[api,pra,amsmath,amssymb,twocolumn]{revtex4-1}

\usepackage{bm}
\usepackage{graphicx}
\usepackage[bookmarks=false]{hyperref}

\def\dd{\mathrm{d}}
\def\ee{\mathrm{e}}
\def\ii{\mathrm{i}}

\def\Re{\mathrm{Re}}
\def\Im{\mathrm{Im}}

\def\ddt#1{\frac{\partial #1}{\partial t}}
\def\ddtt#1{\frac{\partial^2 #1}{\partial t^2}}

\def\die{\varepsilon}
\def\diez{\varepsilon_0}
\def\diep{\varepsilon_{\text{p}}}
\def\np{n_{\text{p}}}
\def\kp{k_{\text{p}}}
\def\muz{\mu_0}
\def\area{S}
\def\len{L}
\def\vg{v_{\text{g}}}
\def\cavlen{l}
\def\cavlenz{l_0}
\def\kappaRWA{\kappa^{\text{RWA}}}
\def\kappaMBC{\kappa_{\text{MBC}}}

\def\gR{\tilde{r}}
\def\gT{\tilde{t}}
\def\GG{\mathcal{G}}

\def\wT{\omega_{\text{ex}}}
\def\wL{\omega_{\text{exL}}}

\def\wc{\varOmega^{\text{cav}}}
\def\rabi{g}
\def\AA{D}

\def\oHp{\hat{H}_{\text{p}}}
\def\oHSR{\hat{H}_{\text{S-E}}}

\def\oJN{\hat{J}_{\text{N}}}
\def\oA{\hat{A}}
\def\oE{\hat{E}}
\def\oB{\hat{B}}
\def\oD{\hat{D}}

\def\oP{\hat{P}}
\def\oJ{\hat{J}}

\def\oa{\hat{a}}
\def\oad{\hat{a}^{\dagger}}
\def\ob{\hat{b}}
\def\obd{\hat{b}^{\dagger}}
\def\op{\hat{p}}
\def\opd{\hat{p}^{\dagger}}
\def\oain{\hat{a}_{\text{in}}}
\def\oaind{\hat{a}_{\text{in}}^{\dagger}}
\def\oaout{\hat{a}_{\text{out}}}

\def\oalpha{\hat{\alpha}}
\def\oalphad{\hat{\alpha}^{\dagger}}

\begin{document}


\title{System-environment coupling derived by Maxwell's boundary conditions from weak to ultrastrong light-matter coupling regime}

\author{Motoaki Bamba}
\altaffiliation{E-mail: bamba@acty.phys.sci.osaka-u.ac.jp}
\affiliation{Department of Physics, Osaka University, 1-1 Machikaneyama, Toyonaka, Osaka 560-0043, Japan}
\author{Tetsuo Ogawa}
\affiliation{Department of Physics, Osaka University, 1-1 Machikaneyama, Toyonaka, Osaka 560-0043, Japan}
\date{\today}

\date{\today}

\begin{abstract}
In the standard theory of cavity quantum electrodynamics (QED),
coupling between photons inside and outside a cavity
(cavity system and environment)
is given conserving the total number of photons.
However, when the cavity photons (ultrastrongly) interact with atoms or excitations in matters,
the system-environment coupling must be determined from a more fundamental viewpoint.
Based on the Maxwell's boundary conditions
in the QED theory for dielectric media,
we derive the quantum Langevin equation and input-output relation,
in which the total number of polaritons (not photons) inside the cavity
and photons outside is conserved.
\end{abstract}


\maketitle
\section{Introduction}
The dissipation has long been discussed
as an inevitable phenomenon in most systems,
and the light is a typical target in the study of such open systems.
When the light is confined in a cavity consisting of mirrors
or distributed Bragg reflectors, discrete cavity modes are well identified,
while they have finite broadenings due to the loss through the mirrors.
In the standard theory of quantum optics \cite{gardiner04,walls08},
coupling between the cavity modes and external photonic field
is usually supposed as
\begin{equation} \label{eq:HSR-cavity} 
\oHSR^{\text{standard}} = \sum_m \int\dd\omega\
  \ii\hbar \sqrt{\frac{\kappa_m(\omega)}{2\pi}}
  \left[ \oalphad(\omega) \oa_m - \oad_m \oalpha(\omega) \right].
\end{equation}
Here, $\oa_m$ is the annihilation operator of a photon in $m$-th cavity mode,
$\oalpha(\omega)$ is the one outside the cavity with frequency $\omega$,
and $\kappa_m(\omega)$ is the dissipation rate of the $m$-th mode.
This expression has successfully reproduced a variety of experimental results,
even when the cavity photons interact with atoms or excitations in matters.
However, in the ultrastrong light-matter coupling regime, where
the vacuum Rabi splitting $\rabi$ is comparable to or larger than 
the transition frequency $\wT$ of excitation in matter
\cite{Ciuti2005PRB,Ciuti2006PRA,Gunter2009N,Anappara2009PRB,Todorov2009PRL,Todorov2010PRL,Todorov2012PRB,Porer2012PRB,Niemczyk2010NP,Fedorov2010PRL,Forn-Diaz2010PRL,Scalari2012S},
we encounter a problem of the treatment of the system-environment coupling
\cite{Beaudoin2011PRA,Ridolfo2012PRL,Bamba2012DissipationUSC}.
This is because the rotating wave approximation (RWA) cannot be applied
on the light-matter coupling,
and the total number of photons and excitations is no longer conserved.
Then, while the number of photons inside and outside the cavity
is conserved in Eq.~\eqref{eq:HSR-cavity},
we should reconsider the validity of this expression carefully.

In the ultrastrong light-matter coupling regime,
even in the ground state of the coupled (polariton) system,
there are virtual photons represented as a ``squeezed'' vacuum state
\cite{Ciuti2005PRB}.
As pointed out by Glauber and Lewenstein \cite{Glauber1991PRA},
such virtual photons exist even in a simple dielectric medium,
and its ``squeezing'' character is different
from that of the squeezed light in vacuum.
Whereas the electromagnetic fields are certainly sub- and super-fluctuant in dielectrics,
such a ``squeezed'' quantum fluctuation recovers to the one of the coherent or vacuum state
when the fields escape from the dielectrics to the vacuum.
In this way, even in the ultrastrong light-matter coupling regime,
the polaritons simply represent the electromagnetic fields in dielectrics,
and we cannot generate non-classical light outside the cavity
at least in the linear optical process with classical inputs.
Certainly, classical outputs are obtained by classical inputs
at least in the approach of quantum Langevin equations \cite{Glauber1991PRA,Ciuti2006PRA}.

However, when we simply suppose the standard expression \eqref{eq:HSR-cavity},
we encounter a delicate but elemental problem:
Even if the outside is the vacuum (bath at zero temperature),
since the virtual photons inside the cavity can escape to the outside,
the polariton system is inevitably excited \cite{Bamba2012DissipationUSC}.
While Eq.~\eqref{eq:HSR-cavity} is introduced phenomenologically in some cases,
its validity can be justified based on a fundamental framework
at least when the cavity is empty and its quality factor is high (good cavity limit)
\cite{Knoll1991PRA,gruner96mar,Dutra2000JOB,Dutra2000PRA,Dalton2001PRA,Hackenbroich2002PRL,Viviescas2003PRA,Khanbekyan2005PRA}.
However, when the cavity is not empty and cavity photons interact with matter,
it is still not clear whether Eq.~\eqref{eq:HSR-cavity} maintains
and the system is really excited by the vacuum or not.
In order to check it, we must derive the system-environment coupling
based on the fundamental framework
under self-consistently considering the light-matter coupling inside the cavity.
In this paper, we simply suppose a dielectric cavity embedding a medium
involving bosonic excitations with single excitation frequency.
In such dielectric system,
the system-environment coupling can be derived
based on the quantum electrodynamics (QED) theory for dielectric media
\cite{huttner92,gruner96mar,gruner96aug,matloob96,knoll01,khanbekyan03,suttorp04,Khanbekyan2005PRA},
and it is determined basically by the Maxwell's boundary conditions.
In the following sections, we show that Eq.~\eqref{eq:HSR-cavity} is not correct in general
when the cavity photons interact with matter.
Instead, we derive another expression of system-environment coupling,
in which the total number of polaritons (not photons) inside the cavity
and photons outside is conserved.

This paper consists as follows. 
In Sec.~\ref{sec:polariton},
we first discuss an homogeneous polariton system without any loss.
In Sec.~\ref{sec:QED}, we show a brief review of the QED theory
for dielectrics.
The cavity structure is introduced in Sec.~\ref{sec:SEC},
and the system-environment coupling is derived.
In Sec.~\ref{sec:comparison},
dissipation rates of polaritons derived in the present work
is quantitatively compared with
the ones obtained from the standard expression \eqref{eq:HSR-cavity}.
They show qualitatively different behaviors in the ultrastrong light-matter
coupling regime.
Some other prospects are discussed in Sec.~\ref{sec:discussion},
and the summary is in Sec.~\ref{sec:summary}.
App.~\ref{app:diag_polariton} shows
a detailed calculation of diagonalizing the polariton system,
and the quantum fluctuation of electromagnetic fields in the medium
is discussed in App.~\ref{app:fluctuation}.
The equivalence between the approach with Maxwell's boundary conditions
and the one with Green's function
is shown in App.~\ref{app:Green}.

\section{Polaritons in homogeneous medium} \label{sec:polariton}
First of all, we consider a loss-less homogeneous dielectric medium,
in which photons interact with infinite-mass excitations
\cite{hopfield58,Ciuti2005PRB}.
The Hamiltonian is represented as
\begin{align} \label{eq:oH_polariton} 
\oHp^{\text{bulk}} & = \sum_{k=-\infty}^{\infty} \left\{
  \hbar c |k| \oad_k \oa_k + \hbar\wT \obd_k \ob_k
\right. \nonumber \\ & \quad
+ \ii\hbar\rabi_k(\oa_k + \oad_{-k})(\ob_{-k}-\obd_{k})
\nonumber \\ & \quad \left.
+ \hbar\AA_k(\oa_k + \oad_{-k})(\oa_{-k} + \oad_{k})
\right\}.
\end{align}
Here, $\oa_k$ and $\ob_k$ are annihilation operators
of a photon and a bosonic excitation with wavenumber $k$ in $z$ direction, respectively,
and satisfy $[\oa_k, \oad_{k'}] = [\ob_k, \obd_{k'}] = \delta_{k,k'}$.
Using the photon operator, the vector potential is represented as
\begin{equation} \label{eq:oA_oa} 
\oA(z)
= \sum_{k=-\infty}^{\infty} \sqrt{\frac{\hbar}{2\diez c|k|\area\len}}
    \left( \oa_k + \oad_{-k} \right) \ee^{\ii kz},
\end{equation}
where $c$ is the speed of light in vacuum,
$\diez$ is the vacuum permittivity, 
$\area$ is the area in $x-y$ plane,
and $\len$ is the length in $z$ direction.
$\wT$ is the frequency of excitations,
$\rabi_k$ is the light-matter coupling strength,
and the coefficient of the last term is $\AA_k = {\rabi_k}^2 / \wT$.
Introducing the annihilation operators
of lower and upper ($j = L$ and $U$) polaritons as
\begin{equation} \label{eq:op_jk} 
\op_{j,k} = w_{jk} \oa_k + x_{jk} \ob_k + y_{jk} \oad_{-k} + z_{jk} \obd_{-k},
\end{equation}
we can diagonalize Eq.~\eqref{eq:oH_polariton}
\cite{hopfield58,Ciuti2005PRB}:
\begin{equation}
\oHp^{\text{bulk}} = \sum_{j=L,U}\sum_{k=-\infty}^{\infty} \hbar\omega_{j,k} \opd_{j,k}\op_{j,k}
+ \text{const}.
\end{equation}
The eigen-frequencies $\omega_{jk}$
and coefficients $\{w_{jk}, x_{jk}, y_{jk}, z_{jk}\}$
are determined by $[\op_{jk}, \oHp] = \hbar\omega_{jk}\op_{jk}$
and $[\op_{jk},\opd_{j'k'}] = \delta_{jj'}\delta_{k,k'}$,
and $\omega = \omega_{jk}$ satisfies
\begin{equation} \label{eq:diep} 
\frac{c^2k^2}{\omega^2} = \diep(\omega)
= 1 + \frac{4\pi\beta{\wT}^2}{\wT{}^2 - (\omega+\ii0^+)^2}.
\end{equation}
Here, $\diep(\omega)$ is the dielectric function of the polariton medium
\cite{hopfield58,matloob96},
and we suppose the coefficient 
$4\pi\beta = 4c|k|{\rabi_k}^2 / {\wT}^3$
does not depend on $k$ for simplicity.
Basically, the light-matter coupling in the dielectric medium
can be described through this dielectric function $\diep(\omega)$.
As discussed in detail in App.~\ref{app:diag_polariton},
by using the polariton operator, the positive-frequency component
(corresponding to annihilation operators)
of vector potential \eqref{eq:oA_oa} is expressed as
\begin{equation} \label{eq:oA_op} 
\oA^+(z)
= \sum_{j=L,U}\sum_{k=-\infty}^{\infty}
  \sqrt{\frac{\hbar \vg(\omega_{jk})}{2\diez c \omega_{jk} \np(\omega_{jk})\area\len}}
  \op_{jk} \ee^{\ii kz}.
\end{equation}
Here, $\np(\omega) = \sqrt{\diep(\omega)}$ is the refractive index, and
\begin{equation}
\vg(\omega) = \frac{\partial\omega}{\partial k}
= \frac{c}{\np(\omega) + \omega(\partial \np/\partial \omega)}
\end{equation}
is the group velocity.
In this way, polaritons represent the eigen-states
of the electromagnetic fields in dielectrics
even in the ultrastrong light-matter coupling regime $4\pi\beta \gtrsim 1$
\cite{Note1}.
As discussed in Ref.~\cite{Glauber1991PRA},
the quantum fluctuation of the electromagnetic fields
is modulated in dielectrics, which can also be verified
from Eq.~\eqref{eq:oA_op} as discussed in App.~\ref{app:fluctuation}.

\section{QED theory for dielectrics} \label{sec:QED}
Next, in order to introduce boundaries of the polariton system,
we employ the QED theory for inhomogeneous media
\cite{gruner96aug,matloob96,knoll01,khanbekyan03,suttorp04,Khanbekyan2005PRA}.
Whereas this theory basically equivalent to the formalism of Green's functions
in absorptive dielectric media \cite{abrikosov75ch6},
it has been developed with compatibility
to the classical electrodynamics (Maxwell's equations)
and to the fluctuation-dissipation theorem.
Here, we simply consider an one-dimensional (1D) system
with dielectric function $\die(z,\omega)$
depending on position $z$ and frequency $\omega$
satisfying the Kramers-Kronig relation,
and the electric and magnetic fields are in the $x-y$ plane.
The positive-frequency component $\oA^+(z,\omega)$ of vector potential
in this system obeys
\begin{equation} \label{eq:Maxwell_wave_A} 
\left[ \frac{\partial^2}{\partial z^2} + \frac{\omega^2}{c^2}\die(z,\omega) \right]
\oA^+(z,\omega) = - \muz \oJN(z,\omega).
\end{equation}
This has exactly the same form as the wave equation derived from the Maxwell equations.
Here, $\muz$ is the vacuum permeability,
and the quantum fluctuation of the electromagnetic fields
is described by the noise current density operator $\oJN(z,\omega)$ satisfying
\begin{equation} \label{eq:[JN,JN]} 
[\oJN(z,\omega), \oJN(z',\omega')^{\dagger}]
= \delta(\omega-\omega') \delta(z-z') \frac{\diez\hbar\omega^2}{\pi\area}\Im[\die(z,\omega)].
\end{equation}
These two equations \eqref{eq:Maxwell_wave_A} and \eqref{eq:[JN,JN]} are derived for general inhomogeneous systems (without optical nonlinearity)
starting from a composite system
consisting of the radiation field, polarizable bosonic excitations,
and bath connecting to the excitations \cite{knoll01,suttorp04}.
The coupling between the radiation field and polarizable excitations
(light-matter coupling) in the dielectric media
can be described through the dielectric function $\die(z,\omega)$,
which generally depends on position $z$.
The noise current density $\oJN(z,\omega)$ corresponds
to the fluctuation operator in the formalism of quantum Langevin equations.
The strength of the fluctuation depends on the absorption $\Im[\die(z,\omega)]$
in the dielectric medium,
and the dynamics of the electromagnetic fields certainly
obeys the fluctuation-dissipation theorem \cite{knoll01}.

The electric and magnetic fields are represented as
$\oE(z,t) = -(\partial/\partial t)\oA(z,t)$
and $\oB(z,t) = (\partial/\partial z)\oA(z,t)$, respectively,
and their positive-frequency components satisfy the Maxwell's equations
\begin{subequations}
\begin{align}
\frac{\partial}{\partial z}\oE^+(z,\omega) & = \ii\omega\oB^+(z,\omega),
\label{eq:kE=B} \\ 
- \frac{1}{\muz}\frac{\partial}{\partial z}\oB^+(z,\omega) & = -\ii\omega\oD^+(z,\omega).
\label{eq:kB=D} 
\end{align}
\end{subequations}
The latter equation is equivalent with Eq.~\eqref{eq:Maxwell_wave_A},
and the displacement field includes the noise current density as
$\oD^+(z,\omega) = \diez\die(z,\omega)\oE^+(z,\omega) + (\ii/\omega)\oJN(z,\omega)$.
Based on this formalism \cite{knoll01,gruner96aug,matloob96,khanbekyan03,suttorp04,Khanbekyan2005PRA},
the positive- and negative-frequency components never mix with each other
at least in the linear optical process.
Then, for polariton system confined in an optical cavity,
we can imagine that the annihilation operator $\op_j$ of polariton
couples with $\oalpha(\omega)$ of photon outside the cavity,
and they never couple with creation operators [$\opd_j$ and $\oalphad(\omega)$]
in the linear optical process.
In order to confirm it,
next we explicitly consider a cavity structure,
and derive the quantum Langevin equation
and input-output relation from the above Maxwell's equations.

\section{System-environment coupling} \label{sec:SEC}
When we first quantize the radiation field in a box (cavity) without any loss,
the system-environment coupling is usually introduced phenomenologically
under the hypothesis that the photon number should be conserved inside and outside a cavity
\cite{gardiner04,walls08}.
However, when a cavity has a loss through its mirrors,
the radiation field is in principle continuously spread inside and also outside the cavity.
The density of states is localized at resonance frequencies
in the good cavity case (the loss rate is smaller than the frequency spacing
of the cavity modes), and then we can well identify the cavity modes,
which are quantized obeying the standard procedure for the loss-less box.
In order to derive rigorously the system-environment coupling Hamiltonian,
we must start from a fundamental principle, which depends on the mechanism of confinement and loss of the field. In the case of the electromagnetic fields, it is the reflectivity of mirrors or more fundamentally the Maxwell's boundary conditions, which gives the (Fresnel's) reflection coefficients.

The system-environment coupling can be derived by supposing reflectivities of cavity mirrors (e.g., Ref.~\cite{Dutra2000JOB}). However, it is not reliable when cavity photons interact with matters, because the reflectivity is in principle modified by the change of refractive index inside the cavity. Instead, the Maxwell's boundary conditions have been used as a more reliable principle connecting the cavity system and the outside
\cite{Knoll1991PRA,gruner96mar,Dutra2000PRA,Dalton2001PRA,Hackenbroich2002PRL,Viviescas2003PRA,Khanbekyan2005PRA} in both classical and quantum electrodynamics. In these works, Eq.~\eqref{eq:HSR-cavity} is certainly obtained for empty cavities with high quality factor. However, its validity has been discussed only in the weak and normally strong light-matter coupling regimes. In order to check the validity of Eq.~\eqref{eq:HSR-cavity} in the ultrastrong light-matter coupling regime, we must derive the system-environment coupling
with self-consistently considering the light-matter coupling inside the cavity.

\begin{figure}[tbp]
\begin{center}
\includegraphics[width=.5\textwidth]{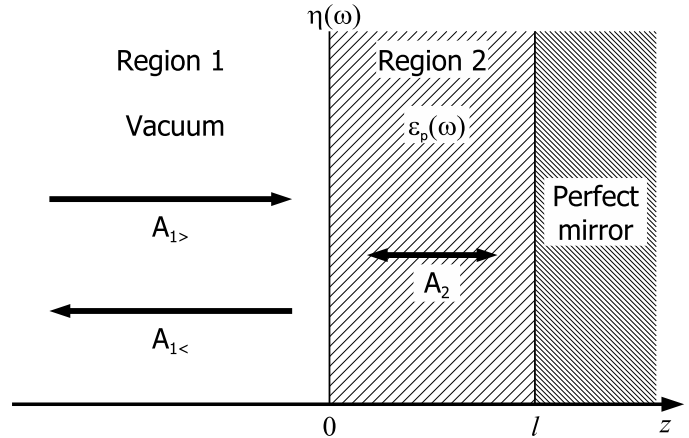}
\end{center}
\caption{Sketch of the considered one-dimensional system with dielectric function
in Eq.~\eqref{eq:die}.}
\label{fig:1}
\end{figure}
As discussed in Ref.~\cite{Lang1973PRA},
we consider a cavity system shown in Fig.~\ref{fig:1}.
The dielectric function is given as
\begin{equation} \label{eq:die} 
\die(z,\omega) = \eta(\omega)\delta(z) + 
\begin{cases}
1 & z < 0 \\
\diep(\omega) & 0 < z < \cavlen
\end{cases}
\end{equation}
There is a perfect mirror at $z = \cavlen$,
and the other mirror is placed at $z = 0$.
$\eta(\omega)$ determines the transparency between the cavity and the outside.
Ref.~\cite{Lang1973PRA} discusses an empty cavity [$\die(0<z<\cavlen,\omega) = 1$],
and the standard system-environment coupling \eqref{eq:HSR-cavity}
can be certainly obtained based on the QED theory for dielectrics \cite{gruner96mar,Khanbekyan2005PRA}.
On the other hand, here the cavity embeds the dielectric medium
described by Hamiltonian \eqref{eq:oH_polariton},
i.e., the dielectric function is described as $\diep(\omega)$ in Eq.~\eqref{eq:diep}
for region 2 ($0 < z < \cavlen$).
By supposing this position-dependent dielectric function \eqref{eq:die},
we can describe the light-matter coupling inside the cavity
and the loss of the electromagnetic fields through the mirror on an equal footing.
In this formalism, we need not explicitly consider the light-matter coupling
as in Eq.~\eqref{eq:oH_polariton},
but it is reflected through the dielectric function $\diep(\omega)$.

For the solution to Eq.~\eqref{eq:Maxwell_wave_A},
we suppose the vector potential in the two regions
as depicted in Fig.~\ref{fig:1}:
\begin{subequations}
\begin{align}
\oA_1^+(z,\omega) & = \oA_{1>}^+(\omega) \ee^{\ii(\omega/c)z}
  + \oA_{1<}^+(\omega) \ee^{-\ii(\omega/c)z}, \\
\oA_2^+(z,\omega) & = \oA_{2}^+(\omega) \sin[\kp(\omega)(\cavlen-z)],
\end{align}
\end{subequations}
where $\kp(\omega) = \np(\omega) \omega / c$.
Since the electric field is completely zero at the boundary with the perfect
mirror, the intra-cavity mode has no amplitude at $z = \cavlen$.
Here, as discussed in Ref.~\cite{knoll01,gruner96aug,matloob96,khanbekyan03,Khanbekyan2005PRA}
and also in App.~\ref{app:Green},
the incoming field $\oA_{1>}$ can be simply derived as
\begin{equation} \label{eq:oA_1>} 
\oA_{1>}^+(\omega)
= \sqrt{\frac{\hbar}{4\pi\diez c\omega \area}} \oa_>(\omega),
\end{equation}
where $\oa_>(\omega)$ is defined as
\begin{equation}
\oa_{>}(\omega)
= \ii \sqrt{\frac{\pi\muz c\area}{\hbar\omega}}
  \int_{-\infty}^0\dd z\ \ee^{-\ii (\omega/c) z} \oJN(z,\omega).
\end{equation}
From Eq.~\eqref{eq:[JN,JN]}, this operator satisfies
$[\oa_>(\omega), \oad_>(\omega')] = \delta(\omega-\omega')$,
and it corresponds to the annihilation operator
of an incoming photon.

Next, we consider boundary conditions determining $\oA_{1<}^+(\omega)$ and $\oA_2^+(\omega)$.
From the continuous condition at $z = 0$
derived from Eq.~\eqref{eq:kE=B}, we get
\begin{subequations} \label{eq:MBC_cavity} 
\begin{equation} \label{eq:MBC_E_cavity} 
\oA_{1>}^+(\omega) + \oA_{1<}^+(\omega)
= \oA_{2}^+(\omega) \sin[\kp(\omega)\cavlen].
\end{equation}
From the integral form of Eq.~\eqref{eq:kB=D}, we also get
\begin{align}
& [\oA_{1>}^+(\omega) - \oA_{1<}^+(\omega)]
- \ii \np(\omega) \oA_{2}^+(\omega) \cos[\kp(\omega)\cavlen] \nonumber \\
& \quad = - \ii\varLambda(\omega) \oA_{2}^+(\omega) \sin[\kp(\omega)\cavlen],
\label{eq:MBC_B_cavity} 
\end{align}
\end{subequations}
where $\varLambda(\omega) = \eta(\omega) \omega/c$.
Solving these Maxwell's boundary conditions, we get
\begin{equation} \label{eq:E2_cav} 
\oA_{2}^+(\omega) = 
\frac{2\oA_{1>}^+(\omega)}
{[1-\ii\varLambda(\omega)]\sin[\kp(\omega)\cavlen]+\ii\np(\omega)\cos[\kp(\omega)\cavlen]},
\end{equation}
and $\oA_{1<}^+(\omega)$ is expressed by Eq.~\eqref{eq:MBC_E_cavity}.
They are also derived from Eq.~\eqref{eq:Maxwell_wave_A}
by using the Green's function as discussed in App.~\ref{app:Green}.
As seen in Eq.~\eqref{eq:E2_cav},
we can find that resonances are obtained at $\omega = \varOmega_{\lambda}$
satisfying
\begin{equation} \label{eq:resonance} 
\tan[\np(\varOmega_{\lambda}) \varOmega_{\lambda}\cavlen / c]
= \np(\varOmega_{\lambda}) / \varLambda(\varOmega_{\lambda})
\end{equation}
and the frequency broadening is proportional to $\varLambda(\varOmega_{\lambda})^{-2}$.
Here, as discussed in Ref.~\cite{Lang1973PRA},
we consider the good cavity limit $\varLambda(\varOmega_{\lambda}) \gg \np(\varOmega_{\lambda})$,
and we suppose that $\varLambda(\omega)$ varies only slightly
around $\omega = \varOmega_{\lambda}$ for simplicity.
Under these assumptions, around the resonance $\omega \sim \varOmega_{\lambda}$,
Eqs.~\eqref{eq:E2_cav} and \eqref{eq:MBC_E_cavity} are approximately rewritten as
\begin{subequations} \label{eq:at_resonance_cavity} 
\begin{equation}
A_2^+(\omega) = \sqrt{\frac{2\vg(\varOmega_{\lambda})}{\np(\varOmega_{\lambda})\cavlen}}
\frac{\ii\sqrt{\kappaMBC(\varOmega_{\lambda})}}{\omega-\varOmega_{\lambda}+\ii\kappaMBC(\varOmega_{\lambda})/2} A_{1>}^+(\omega),
\end{equation}
\begin{equation}
A_{1>}^+(\omega) + A_{1<}^+(\omega)
= \sqrt{\frac{\np(\varOmega_{\lambda})\cavlen}{2\vg(\varOmega_{\lambda})}}
    \sqrt{\kappaMBC(\varOmega_{\lambda})} A_2^+(\omega),
\end{equation}
\end{subequations}
where the dissipation rate $\kappaMBC(\omega)$ is defined as
\begin{equation} \label{eq:kappa} 
\kappaMBC(\omega) = \frac{2\np(\omega)\vg(\omega)}{\varLambda(\omega)^2\cavlen}.
\end{equation}
In the semi-infinite region 1,
$\oA_{1<}^+(\omega)$ is expressed by
the annihilation operator $\oa_{1<}(\omega)$ of an outgoing photon as
\begin{equation} \label{eq:oA_1<} 
\oA_{1<}^+(\omega)
= \sqrt{\frac{\hbar}{4\pi\diez c\omega\area}} \oa_{1<}(\omega).
\end{equation}
On the other hand, in the finite region 2,
in the similar way for deriving Eq.~\eqref{eq:oA_op},
the expression of $\oA_2^+(z)$ is represented
in the good cavity limit:
\begin{equation} \label{eq:oA_2} 
\oA_2^+(z)
= \sum_{j=L,U} \sum_{m=1}^{\infty}
  \sqrt{\frac{\hbar\vg(\varOmega_{jm})}{\diez c \varOmega_{jm} \np(\varOmega_{jm})\area\cavlen}}
    \op_{jm}
    \sin[k_m(\cavlen-z)],
\end{equation}
where $k_m = m\pi / \cavlen$, $\varOmega_{jm} = \omega_{j,k=k_m}$
and $[\op_{jm},\opd_{j'm'}] = \delta_{j,j'}\delta_{m,m'}$.
The index $j = L,U$ means lower and upper polariton state,
and the previous index is replaced as $\lambda \rightarrow jm$.
Then, from Eqs.~\eqref{eq:oA_1>}, \eqref{eq:oA_1<}, and \eqref{eq:oA_2},
Eqs.~\eqref{eq:at_resonance_cavity} are rewritten
in the good cavity limit as
\begin{subequations} \label{eq:Langevin_inout} 
\begin{equation} \label{eq:Langevin} 
\op_{jm}(\omega) = \frac{\ii\sqrt{\kappaMBC(\varOmega_{jm})}}{\omega-\varOmega_{jm}+\ii\kappaMBC(\varOmega_{jm})/2} \oain(\omega),
\end{equation}
\begin{equation} \label{eq:input-output} 
\oain(\omega) + \oaout(\omega) = \sum_{j=L,U}\sum_{m=1}^{\infty}\sqrt{\kappaMBC(\varOmega_{jm})} \op_{jm}(\omega).
\end{equation}
\end{subequations}
where the input and output operators are defined as
$\oain(\omega) = \oa_{1>}(\omega)/ \sqrt{2\pi}$
and 
$\oaout(\omega) = \oa_{1<}(\omega) / \sqrt{2\pi}$.
Eqs.~\eqref{eq:Langevin_inout} are derived
from Eq.~\eqref{eq:Maxwell_wave_A} by supposing only the good cavity limit.
This fact indicates that
the coupling between the cavity polariton system
and its surrounding is expressed as
\begin{widetext}
\begin{equation} \label{eq:HSR-MBC} 
\oHSR^{\text{MBC}} = \sum_{j=L,U} \sum_{m=1}^{\infty} \int\dd\omega\
  \ii\hbar \sqrt{\frac{\kappaMBC(\varOmega_{jm})}{2\pi}}
  \left[ \oalphad(\omega) \op_{jm} - \opd_{jm} \oalpha(\omega) \right].
\end{equation}
\end{widetext}
Obeying the well-known treatment in quantum optics \cite{gardiner04,walls08},
Eqs.~\eqref{eq:Langevin_inout} are certainly derived from Eq.~\eqref{eq:HSR-MBC}
as the quantum Langevin equation of polariton
and the input-output relation.
Since the annihilation operator $\op_{jm}$ of polaritons
is expressed not only by $\oa$ and $\ob$
but also by the creation ones $\oad$ and $\obd$ as in Eq.~\eqref{eq:op_jk},
the derived system-environment coupling \eqref{eq:HSR-MBC}
is in principle different from the standard one \eqref{eq:HSR-cavity},
which is justified only for empty cavities.
If we suppose Eq.~\eqref{eq:HSR-cavity} as the system-environment coupling,
the creation operator $\oaind$ of input is added in the quantum Langevin equation \eqref{eq:Langevin},
and then the polaritons are excited by the vacuum.
However, since the system-environment coupling is in fact expressed
as Eq.~\eqref{eq:HSR-MBC},
the polaritons are in principle not excited by the vacuum.

\section{Quantitative comparison} \label{sec:comparison}
To check the consistency with the well-known discussions \cite{gardiner04,walls08},
we suppose the simplified case where
the resonance frequency of the lowest bare cavity mode is tuned to $\wT$,
and we calculate the dissipation rates by Eq.~\eqref{eq:kappa}.
For simplicity, we suppose $\varLambda(\omega)$ is constant
in the frequency range of interest.
In the good cavity limit $\varLambda \gg 1$,
the cavity length is determined
as $\cavlenz \simeq \pi c / \wT$ satisfying $\tan(\wT\cavlen/c) \ll 1$.
Then, from Eq.~\eqref{eq:kappa},
the dissipation rate of bare cavity mode is obtained as
\begin{equation}
\kappa_0 = 2c/\varLambda^2\cavlenz.
\end{equation}
From Eq.~\eqref{eq:resonance},
the frequencies of lowest upper and lower polariton modes
($m = 1$ is omitted in the followings) are determined by
$n(\varOmega_{L,U}) = \wT / \varOmega_{L,U}$,
and Eq.~\eqref{eq:kappa} is rewritten as
\begin{equation} \label{eq:kappa_simple} 
\kappaMBC(\varOmega_{L,U}) \simeq \frac{\kappa_0}{1 + (\varOmega_{L,U}/\wT)^2}.
\end{equation}
If the light-matter coupling is not so strong $4\pi\beta \ll 1$,
we get $\varOmega_{L/U} \simeq \wT \pm \rabi$,
and $\kappaMBC(\varOmega_{L,U})$ is approximately a half of $\kappa_0$.
Then, in the weak and normally strong coupling regimes,
the resonance frequencies of cavity polaritons
and their dissipation rates certainly agree with
the well-known ones.

Next, we compare our results with those derived by supposing
the standard expression \eqref{eq:HSR-cavity}
especially in the ultrastrong light-matter coupling regime.
Whereas the polaritons are excited in general by supposing Eq.~\eqref{eq:HSR-cavity},
it is also avoidable by employing the procedure
discussed in Ref.~\cite{Beaudoin2011PRA,Ridolfo2012PRL}.
In this procedure, the system-environment coupling \eqref{eq:HSR-cavity}
is determined by supposing an empty cavity
or is simply introduced phenomenologically,
and $\kappa_m$ is the dissipation rate of bare cavity photons in $m$-th mode.
The light-matter coupling is additionally introduced as follows.
In the basis of the well-defined cavity modes
determined for the empty cavity,
compared with Eq.~\eqref{eq:oH_polariton},
the Hamiltonian inside the cavity is represented as
\begin{align} \label{eq:oHp-discrete} 
\oHp^{\text{discrete}} & = \sum_{m} \left\{
  \hbar \wc_m \oad_m \oa_m + \hbar\wT \obd_m \ob_m
\right. \nonumber \\ & \quad
+ \ii\hbar\rabi_m(\oa_m + \oad_{m})(\ob_{m}-\obd_{m})
\nonumber \\ & \quad \left.
+ \hbar\AA_m(\oa_m + \oad_{m})(\oa_{m} + \oad_{m})
\right\}.
\end{align}
Here, $\wc_m$ is the resonance frequency of the $m$-th cavity mode,
and $\oa_m$ and $\ob_m$ are the annihilation operators of cavity photon
and excitation in $m$-the mode, respectively.
Since we suppose that the infinite-mass excitations,
a cavity mode couples with an excitation mode with the same wavefunction
inside the cavity.
Then, we can expand the excitation modes in the same orthogonal basis
as the cavity ones.
The vacuum Rabi splitting $\rabi_m$ and $\AA_m$ are determined
depending on the wavefunction of the $m$-th mode.
The Hamiltonian \eqref{eq:oHp-discrete} also can be diagonalized
by the polariton operator \cite{Ciuti2005PRB}:
\begin{equation}
\op_{jm} = w_{jm} \oa_m + x_{jm} \ob_m + y_{jm} \oad_{m} + z_{jm} \obd_{m}.
\end{equation}
Inversely, the annihilation operator $\oa_m$ of cavity photon in $m$-th mode
is represented by the polariton operator $\op_{jm}$ as
\begin{equation}
\oa_m = \sum_{j=L,U} \left( w_{jm}^* \op_{jm} - y_{jm}\opd_{jm} \right).
\end{equation}
Therefore, the standard expression \eqref{eq:HSR-cavity} is rewritten as
\begin{widetext}
\begin{equation} \label{eq:HSR-cavity-p} 
\oHSR^{\text{standard}} = \sum_{j=L,U} \sum_m \int\dd\omega\
  \ii\hbar \sqrt{\frac{\kappa_m(\omega)}{2\pi}}
  \left[ \oalphad(\omega) \left( w_{jm}^* \op_{jm} - y_{jm}\opd_{jm} \right)
- \left( w_{jm} \opd_{jm} - y_{jm}^*\op_{jm} \right)\oalpha(\omega) \right].
\end{equation}
This expression is clearly different from our expression \eqref{eq:HSR-MBC},
and the polaritons are in general excited by the vacuum
because of the existence of the counter-rotating terms
[$\op_{jm}\oalpha(\omega)$ and $\oalphad(\omega)\opd_{jm}$].
However, the degree of excitation can be decreased
with the decrease of the dissipation rate $\kappa_m$ \cite{Bamba2012DissipationUSC}.
Then, if $\kappa_m$ is small enough compared to the typical frequencies
of cavity polariton system (such as $\wc_m$, $\wT$, and $\rabi_m$),
we can justify the RWA on the system-environment coupling
with respect to the eigen-states (polariton operators)
\cite{Beaudoin2011PRA,Ridolfo2012PRL}.
By neglecting the counter-rotating terms,
Eq.~\eqref{eq:HSR-cavity-p} is approximated as
\begin{equation} \label{eq:HSR-RWA} 
\oHSR^{\text{standard}} \simeq \sum_{j=L,U} \sum_m \int\dd\omega\
  \ii\hbar \sqrt{\frac{\kappa_m(\omega)}{2\pi}}
  \left[ w_{jm}^* \oalphad(\omega) \op_{jm}
       - w_{jm} \opd_{jm} \oalpha(\omega) \right].
\end{equation}
\end{widetext}
This expression certainly has the same form as Eq.~\eqref{eq:HSR-MBC}
derived in this paper.
The quantum Langevin equation and input-output relation
are also derived in the same form as Eqs.~\eqref{eq:Langevin_inout}.
Further, obeying the well-known treatment \cite{gardiner04,walls08},
master equations and photon counting can also be considered.
\cite{Beaudoin2011PRA,Ridolfo2012PRL}.
However, as seen in Eq.~\eqref{eq:HSR-RWA},
the dissipation rate is obtained as
\begin{equation} \label{eq:kappaRWA} 
\kappaRWA_{jm} = |w_{jm}|^2\kappa_m.
\end{equation}
Here, $\kappa_m$ is the dissipation rate of bare cavity mode
determined for the empty cavity as noted above,
and $w_{jm}$ is determined by the Hamiltonian
\eqref{eq:oHp-discrete}.
We have to check whether Eq.~\eqref{eq:kappaRWA}
is really equivalent with Eq.~\eqref{eq:kappa},
which is derived under self-consistently considering the light-matter coupling
and the system-environment one.

\begin{figure}[tbp]
\includegraphics[width=.5\textwidth]{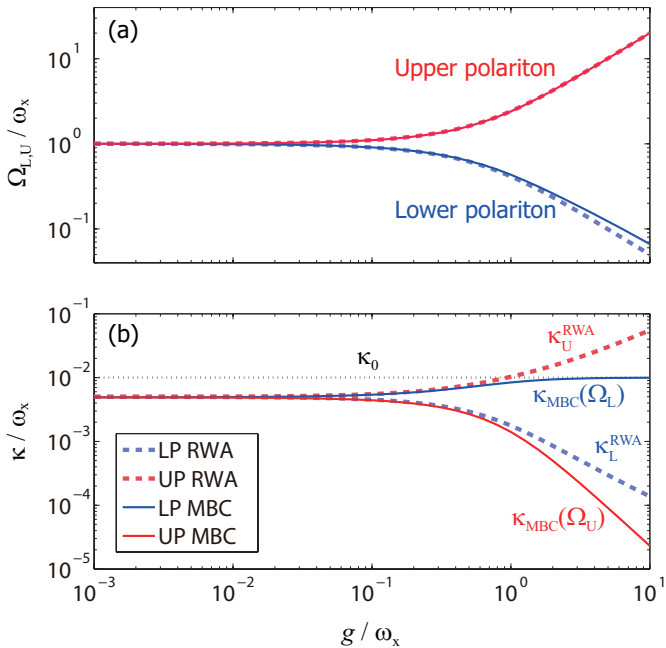}
\caption{(a) Frequencies of lower and upper polaritons (LP and UP)
are plotted versus $\rabi/\wT$.
Dashed curves are calculated by Hamiltonian \eqref{eq:oHp-discrete},
while solid curves are obtained by Eq.~\eqref{eq:resonance}.
(b) Dissipation rates of the polaritons are plotted.
Dashed curves are calculated from Eq.~\eqref{eq:kappaRWA}
with $\kappa_0 / \wT = 10^{-2}$.
Solid curves are obtained from Eq.~\eqref{eq:kappa}
by supposing $\varLambda = 7.822$
and $\cavlen$ tuned for satisfying $\tan(\wT\cavlen/c) = 1 / \varLambda$.
Dotted line represents $\kappa_0/\wT$.}
\label{fig:2}
\end{figure}
As a demonstration, we consider again the simple case
where the lowest bare cavity frequency $\wc_{m=1} = \wc_0$ is tuned to $\wT$.
We focus on only this lowest mode
and index $m$ is omitted in the following discussion.
In Fig.~\ref{fig:2}, we plot (a) frequencies $\varOmega_{L,U}$ of lower and upper
polaritons (LP and UP) and (b) their dissipation rates
as functions of $\rabi_{m=1} / \wT = \sqrt{4\pi\beta}/2$.
For dashed curves, $\varOmega_{L,U}$ are calculated
by diagonalizing Hamiltonian \eqref{eq:oHp-discrete}
of the well-defined discrete mode,
and the dissipation rates are calculated by Eq.~\eqref{eq:kappaRWA}
with supposing the dissipation rate
$\kappa_{m=1}(\omega) = \kappa_0 = 10^{-2}\wT$ of the bare cavity mode.
For solid curves, $\varOmega_{L,U}$ and $\kappaMBC(\varOmega_{L,U})$
are determined by Eqs.~\eqref{eq:resonance} and \eqref{eq:kappa},
respectively, which are obtained by the Maxwell's boundary conditions.
The mirror transparency is assumed as $\varLambda = 7.822$
corresponding to $\kappa_0 / \wT = 10^{-2}$ for the empty case.

As seen in Fig.~\ref{fig:2},
the two approaches give almost the same results
in the weak and normally strong light-matter coupling regimes $\rabi \ll \wT$,
where the frequencies are given as $\varOmega_{L,U} \simeq \wT \pm \rabi$
and the dissipation rates are $\kappa_0 / 2$.
In the ultrastrong light-matter coupling regime $\rabi \gtrsim \wT$,
concerning the resonance frequencies $\varOmega_{L,U}$,
the two approaches agree well with each other at least qualitatively,
while there is a discrepancy for the lower polariton frequency $\varOmega_{L}$
(this is because $\varOmega_L$ closes to the dissipation rate $\kappa_0$).
However, as seen in Fig.~\ref{fig:2}(b), the dissipation rates
obtained by the two approaches show qualitatively different behaviors.
The solid curves obtained by the Maxwell's boundary conditions can be well fitted by Eq.~\eqref{eq:kappa_simple}.
The effective cavity length $\cavlen/\np(\omega)$ increases for upper polariton,
while the group velocity $\vg(\omega)$ remains almost constant.
Then, its dissipation rate (inverse of round trip time) is decreased
obeying Eq.~\eqref{eq:kappa}.
On the other hand, the changes of $\cavlen/\np(\omega)$ and $\vg(\omega)$
are canceled with each other for lower polariton,
then the dissipation rate becomes close to $\kappa_0$.
However, the dashed curves $\kappaRWA_{L,U}$,
which are calculated by supposing the standard system-environment coupling
\eqref{eq:HSR-cavity}, show opposite behavior:
The dissipation rate of lower (upper) polariton mode
is decreased (increased) with the increase of light-matter interaction $\rabi$.
Even if we suppose the expression
such as $(\oa\pm\oad)(\oalpha\mp\oalphad)$
instead of Eq.~\eqref{eq:HSR-cavity} \cite{Beaudoin2011PRA,Ridolfo2012PRL},
$\kappaRWA_{L,U}$ still show qualitatively different behaviors compared with the solid curves.
The failure of Eq.~\eqref{eq:kappaRWA} is simply
because it is derived from the standard system-environment coupling
\eqref{eq:HSR-cavity}, which is justified basically for empty cavities.
In this way, in the ultrastrong light-matter coupling regime,
the standard expression \eqref{eq:HSR-cavity} give incorrect dissipation rates
compared with Eq.~\eqref{eq:kappa} derived by the Maxwell's boundary conditions.

\section{Discussion} \label{sec:discussion}
Whereas the number of photons is conserved
in the standard expression \eqref{eq:HSR-cavity},
this conservation itself is not a principle
for connecting the cavity system and its surroundings.
In principle, the system-environment coupling is determined
depending on the mechanism of confinement and loss of the fields
inside the cavity.
For the electromagnetic fields,
it is determined obeying the Maxwell's equations
or the Maxwell's boundary conditions.
In this spirit, the system-environment coupling has been discussed
for empty cavities.
Refs.~\cite{gruner96mar,Khanbekyan2005PRA} take the similar approach as ours,
the approach based on the universe modes are discussed in Refs.~\cite{Knoll1991PRA,Dutra2000JOB,Dutra2000PRA,Dalton2001PRA},
and Refs.~\cite{Hackenbroich2002PRL,Viviescas2003PRA}
employ the Freshbach's projector approach.
In the present work,
the standard expression \eqref{eq:HSR-cavity}
is justified approximately for weak and normally strong
light-matter coupling regimes,
although we must consider the correct one \eqref{eq:HSR-MBC}
in the ultrastrong light-matter coupling regime.
While empty cavities are discussed in the previous works,
our result is certainly consistent with them.
Although the counter-rotating terms such as $\alpha(\omega)\op_j$
and $\opd_j\oalphad(\omega)$ might appear in general
due to the overlap between the cavity modes
\cite{Dutra2000JOB,Hackenbroich2002PRL},
they are negligible thanks to the quality of the cavity,
not to the RWA on system-environment coupling
as introduced in the procedure discussed in Refs.~\cite{Beaudoin2011PRA,Ridolfo2012PRL}.

As seen in Fig.~\ref{fig:2}(b),
the dissipation rates of the cavity polaritons
derived from the standard expression \eqref{eq:HSR-cavity}
show the qualitatively different behavior
compared with Eq.~\eqref{eq:kappa}
derived by the Maxwell's boundary conditions,
whereas they agree well with each other in the weak and normally strong light-matter coupling regimes.
This can be understood in the sense of perturbation.
If the light-matter coupling is stronger than the coupling with environment,
we should diagonalize the cavity system first,
then the coupling with environment must be determined after.
Then, our expression \eqref{eq:HSR-MBC} is appropriate in general for the entire coupling regimes.
Although the standard one \eqref{eq:HSR-cavity} seems better for the weak coupling regime
in the sense of perturbation theory, Eqs.~\eqref{eq:HSR-cavity}
and \eqref{eq:HSR-MBC} are equivalent in that regime,
because there is little virtual photons and excitations (coefficients $y_j$ and $z_j$
are negligible) and the RWA can be applied on the light-matter coupling.
Our result means also that,
when we consider pumping of polaritons through the cavity mirrors,
we should suppose the polaritons (not the cavity photons) are directly pumped by incident light through the mirrors.

If we want to discuss also the bad cavity case,
we have to explicitly consider the Maxwell's boundary conditions \eqref{eq:MBC_cavity},
and the simplified expressions
\eqref{eq:HSR-cavity} and \eqref{eq:HSR-MBC} cannot be used.
Whereas we simply supposed the cavity system depicted in Fig.~\ref{fig:1},
other cavity structures (e.g., distributed Bragg reflectors) can also be considered
based on the QED theory for dielectrics \cite{gruner96aug,matloob96,knoll01,khanbekyan03,suttorp04,Khanbekyan2005PRA}.
If we consider a loss or a gain inside the cavity
by supposing the imaginary part of the dielectric function $\diep(\omega)$,
the Langevin equation \eqref{eq:Langevin}
would have another noise term expressed by $\oJN(0<z<\cavlen,\omega)$.
This work remains for the future.

In the present paper, we supposed the simple dielectric (polariton) medium,
and our results are applicable to the intersubband transition
in semiconductor quantum wells \cite{Gunter2009N,Anappara2009PRB,Todorov2009PRL,Todorov2010PRL,Todorov2012PRB,Porer2012PRB}
and more traditionally optical phonons,
both of which show the ultrastrong light-matter coupling.
On the other hand,
for cavity systems with single atom (nonlinear systems)
and for the Dicke model (the gauge invariance is broken)
\cite{Emary2004PRA,Nataf2010NC,Viehmann2011PRL},
it is still open to dispute whether the system-environment
coupling is generally expressed by the eigen-states of the cavity system
as in Eq.~\eqref{eq:HSR-MBC}.
Also for the superconducting circuits
\cite{Niemczyk2010NP,Fedorov2010PRL,Forn-Diaz2010PRL},
the system-environment coupling
under the ultrastrong coupling between artificial atoms and resonator modes
should be determined by
an appropriate microscopic description \cite{Yurke1984PRA,Blais2004PRA,Devoret2007AP}.
Even for such works, our result indicates a universal policy:
The system-environment coupling must be determined
from a fundamental viewpoint under self-consistently considering
ultrastrong coupling in composite systems.

\section{Summary} \label{sec:summary}
Based on the QED theory for dielectric media
\cite{huttner92,gruner96mar,gruner96aug,matloob96,knoll01,khanbekyan03,suttorp04,Khanbekyan2005PRA},
from the weak to ultrastrong light-matter coupling regime,
the quantum Langevin equation \eqref{eq:Langevin}
of discrete cavity polariton modes
and the input-output relation \eqref{eq:input-output}
are derived in the good cavity limit.
They suggest that Eq.~\eqref{eq:HSR-MBC} is appropriate for the coupling
between the polariton system and the environment
in the good cavity case.
Then, the total number of cavity polaritons and external photons is conserved,
instead of the conservation of photon number.
Although the standard expression \eqref{eq:HSR-cavity} indicates that
an incoming photon creates a cavity photon and then it couples with matter
inside the cavity,
this interpretation is found to be incorrect.
Eq.~\eqref{eq:HSR-MBC} instead suggests that
an incoming photon not only creates the superposition of a photon and an excitation
but also annihilates them from the microscopic viewpoint.
This fact shows another aspect of ultrastrong light-matter coupling
in the study of open systems and for the generation of non-classical states of light.

\begin{acknowledgments}
M.~B.~thanks to Howard Carmichael and Pierre Nataf
for informing important articles.
This work was supported by KAKENHI (No.~20104008 and No.~24-632)
and the JSPS through its FIRST Program.
\end{acknowledgments}

\appendix
\section{Diagonalization of polariton system} \label{app:diag_polariton}
The Hamiltonian of homogeneous polariton system is expressed in Eq.~\eqref{eq:oH_polariton}.
The vector potential and the electric field are expressed by $\oa_k$ as
\begin{subequations}
\begin{align}
\oA_k
& = \sqrt{\frac{\hbar}{2\diez c|k|\area}} ( \oa_k + \oad_{-k} ), \\
\oE_k
& = \ii\sqrt{\frac{\hbar c|k|}{2\diez\area}} ( \oa_k - \oad_{k} ).
\end{align}
The excitonic polarization and current density are written in $k$-space as
\begin{align}
\oP_k
& = \sqrt{\frac{2\pi\beta\diez\hbar\wT}{\area}} ( \ob_k + \obd_{-k} ), \\
\oJ_k
& = (-\ii\wT)\sqrt{\frac{2\pi\beta\diez\hbar\wT}{\area}} ( \ob_k - \obd_{-k} )
- 4\pi\beta\diez{\wT}^2\oA_k.
\end{align}
\end{subequations}
The equations of motion are derived as
\begin{subequations}
\begin{align}
\ddt{}\oa_k & = - \ii c|k|\oa_k + \rabi_k(\ob_k-\obd_{-k}) - \ii2\AA_k(\oa_k+\oad_{-k}), \\
\ddt{}\ob_k & = - \ii\wT\ob_k - \rabi_k(\oa_k+\oad_{-k}).
\end{align}
\end{subequations}
The equations of motion of the macroscopic fields are obtained as
\begin{subequations}
\begin{align}
\ddt{}\oA_k & = - \oE_k, \\
\ddt{}\oE_k & = c^2k^2\oA_k - \frac{1}{\diez}\oJ_k, \\
\ddt{}\oP_k & = \oJ_k, \\
\ddt{}\oJ_k & = -{\wT}^2\oP_k - 4\pi\beta\diez{\wT}^2\ddt{}\oA_k.
\end{align}
\end{subequations}
The first two equations correspond to the Maxwell equations.
Then, the wave equations are obtained as
\begin{subequations}
\begin{align}
k^2\oA_k + \frac{1}{c^2}\ddtt{}\oA_k & = \muz\oJ_k, \\
k^2\oE_k + \frac{1}{c^2}\ddtt{}\oE_k & = -\muz\ddtt{}\oP_k.
\end{align}
\end{subequations}
The equation of motion of the polarization is rewritten as
\begin{equation}
\ddtt{}\oP_k = -{\wT}^2\oP_k + 4\pi\beta\diez{\wT}^2\oE_k.
\end{equation}
Then, by Fourier transforming to the frequency domain,
we get the dielectric function as
\begin{equation}
\diep(\omega) = \frac{c^2k^2}{\omega^2}
= 1 + \frac{4\pi\beta{\wT}^2}{\wT{}^2 - (\omega+\ii0^+)^2}
= \frac{{\wL}^2 - \omega^2}{\wT{}^2 - (\omega+\ii0^+)^2}.
\end{equation}
Here, $\wL$ is the frequency of longitudinal excitation satisfying
\begin{equation}
4\pi\beta = \frac{\wL{}^2}{\wT{}^2}-1.
\end{equation}

The Hamiltonian can be diagonalized by the Bogoliubov transformation.
The annihilation operator of lower ($j = L$) or upper ($j = U$) polariton
is represented as Eq.~\eqref{eq:op_jk}.
Then, the coefficients are determined by the following eigen-value problem:
\begin{widetext}
\begin{equation}
\begin{pmatrix}
c|k|+2\AA_k & -\ii\rabi_k & -2\AA_k & -\ii\rabi_k \\
\ii\rabi_k & \wT & -\ii\rabi_k & 0 \\
2\AA_k & -\ii\rabi_k & -c|k|-2\AA_k & -\ii\rabi_k \\
-\ii\rabi_k & 0 & \ii\rabi_k & -\wT
\end{pmatrix}
\begin{pmatrix} w_{j,k} \\ x_{j,k} \\ y_{j,k} \\ z_{j,k} \end{pmatrix}
= \omega_{j,k} \begin{pmatrix} w_{j,k} \\ x_{j,k} \\ y_{j,k} \\ z_{j,k} \end{pmatrix}.
\end{equation}
The four eigen-states corresponds to $\op_{j,k}$ and $\opd_{j,k}$,
whose eigen-frequencies are respectively $\omega_{j,k}$ and $-\omega_{j,k}$:
\begin{equation}
\omega_{L/U,k} = \frac{\wT}{\sqrt{2}}
\left\{ 1 + 4\pi\beta + \frac{c^2|k|^2}{{\wT}^2}
 \mp \left[ \left(1 + 4\pi\beta + \frac{c^2|k|^2}{{\wT}^2} \right)^2 - \frac{4c^2|k|^2}{{\wT}^2} \right]^{1/2} \right\}^{1/2}.
\end{equation}
The eigen vectors are derived as \cite{hopfield58}
\begin{equation}
\begin{pmatrix} w_{L,k} \\ x_{L,k} \\ y_{L,k} \\ z_{L,k} \end{pmatrix}
= \left\{ \frac{\omega_L}{\wT}\left[ \left(1-\frac{{\omega_L}^2}{{\wT}^2}\right)^2 + 4\pi\beta \right]\right\}^{-1/2}
\begin{pmatrix}
\left[1-\frac{{\omega_L}^2}{{\wT}^2}\right]\frac{\omega_L+c|k|}{2\wT}\sqrt{\frac{\wT}{c|k|}} \\
-\ii\sqrt{\pi\beta}\left(1 + \frac{\omega_L}{\wT}\right) \\
\left[1-\frac{{\omega_L}^2}{{\wT}^2}\right]\frac{\omega_L-c|k|}{2\wT}\sqrt{\frac{\wT}{c|k|}} \\
-\ii\sqrt{\pi\beta}\left(1 - \frac{\omega_L}{\wT}\right)
\end{pmatrix}.
\end{equation}
\begin{equation}
\begin{pmatrix} w_{U,k} \\ x_{U,k} \\ y_{U,k} \\ z_{U,k} \end{pmatrix}
= \left\{ \frac{\omega_U}{\wT}\left[ \left(1-\frac{{\omega_U}^2}{{\wT}^2}\right)^2 + 4\pi\beta \right]\right\}^{-1/2}
\begin{pmatrix}
-\left[1-\frac{{\omega_U}^2}{{\wT}^2}\right]\frac{\omega_U+c|k|}{2\wT}\sqrt{\frac{\wT}{c|k|}} \\
\ii\sqrt{\pi\beta}\left(1 + \frac{\omega_U}{\wT}\right) \\
-\left[1-\frac{{\omega_U}^2}{{\wT}^2}\right]\frac{\omega_U-c|k|}{2\wT}\sqrt{\frac{\wT}{c|k|}} \\
\ii\sqrt{\pi\beta}\left(1 - \frac{\omega_U}{\wT}\right)
\end{pmatrix}.
\end{equation}
\end{widetext}
We have two eigen-frequencies $\omega_{j,k}$ for a given wavenumber $k$.
Inversely, we get one allowed  wavenumber $k(\omega)$ for a given frequency $\omega$.
The relation of them are expressed by the dielectric function as
\begin{subequations}
\begin{align}
\frac{c^2k^2}{{\omega_{j,k}}^2} & = \diep(\omega_{j,k}), \\
\frac{c^2k(\omega)^2}{\omega^2} & = \diep(\omega).
\end{align}
\end{subequations}
The Hamiltonian is rewritten as
\begin{subequations}
\begin{align}
\oHp
& =  \sum_{j=L,U}\sum_{k=-\infty}^{\infty} \hbar\omega_{j,k} \opd_{j,k}\op_{j,k}
+ \text{const.} \\
& =  \int_{0}^{\infty}\dd \omega\ \hbar\omega
     \left[ \opd_>(\omega)\op_>(\omega) + \opd_<(\omega)\op_<(\omega) \right]
+ \text{const.},
\end{align}
\end{subequations}
where the forward and backward polariton operators ($\op_>$ and $\op_<$)
are defined in the frequency domain as
\begin{align}
\op_{\gtrless}(\omega)
& = \sqrt{\frac{\len}{2\pi\vg(\omega)}}
    \left[ \theta(\omega-\wL) \op_{U,\pm k(\omega)}
         + \theta(\omega-\wT) \op_{L,\pm k(\omega)} \right].
\end{align}
We choose the phase difference between lower and upper polaritons
by the phase of the eigen-vectors shown above
for deriving a simple expression in the followings.
The group velocity is expressed as
\begin{equation}
\vg(\omega)
= \frac{c^2k}{\omega} \left( \frac{\omega^2}{{\wT}^2} - 1 \right)^2
  \left[  \left( \frac{\omega^2}{{\wT}^2} - 1 \right)^2 + 4\pi\beta \right]^{-1}.
\end{equation}

By using the polariton operator,
the original photon and excitation operators are expressed as
\begin{subequations}
\begin{align}
\oa_k & = \sum_{j=L,U} ( w_{j,k}^* \op_{j,k} - y_{j,k} \opd_{j,-k} ), \\
\ob_k & = \sum_{j=L,U} ( x_{j,k}^* \op_{j,k} - z_{j,k} \opd_{j,-k} ).
\end{align}
\end{subequations}
Then, the vector potential
at position $z$ is represented as
\begin{widetext}
\begin{subequations}
\begin{align}
\oA(z)
& = \frac{1}{\sqrt{\len}} \sum_{k=-\infty}^{\infty} \ee^{\ii kz} \oA_k \\
& = \sum_{k=-\infty}^{\infty}
    \sqrt{\frac{\hbar}{2\diez c|k|\area\len}} ( \oa_k + \oad_{-k} ) \ee^{\ii kz} \\
& = \sum_{k=-\infty}^{\infty} \sqrt{\frac{\hbar}{2\diez c|k|\area\len}}
    \sum_{j=L,U} \left[ (w_{j,k}^*-y_{j,-k}^*) \op_{j,k}
    + (w_{j,-k}-y_{j,k}) \opd_{j,-k} \right]\ee^{\ii kz}.
\end{align}
\end{subequations}
The positive-frequency and forward component is then written as
\begin{equation}
\oA^+_>(z)
= \sum_{k=0}^{\infty} \sqrt{\frac{\hbar}{2\diez c|k|\area\len}}
  \sum_{j=L,U} (w_{j,k}^*-y_{j,-k}^*) \op_{j,k} \ee^{\ii kz}.
\end{equation}
By rewriting the polariton operator in the frequency domain, we get
\begin{subequations}
\begin{align}
\oA^+_>(z)
& = \sum_{j=L,U} \sum_{k=0}^{\infty} \sqrt{\frac{\hbar}{2\diez c k\area\len}}
    \frac{
    \left|1-\frac{{\omega_{jk}}^2}{{\wT}^2}\right|
    \sqrt{\frac{ck}{\wT}}
    }{
    \sqrt{ \frac{\omega_{jk}}{\wT}\left[
      \left(1-\frac{{\omega_{jk}}^2}{{\wT}^2}\right)^2 + 4\pi\beta
    \right]}
    }
    \op_{j,k}\ee^{\ii kz} \\
& = \sum_{j=L,U}\sum_{k=-\infty}^{\infty}
  \sqrt{\frac{\hbar \vg(\omega_{jk})}{2\diez c^2 k\area\len}}
  \op_{jk} \ee^{\ii kz} \\
& = \int_{0}^{\infty}\dd \omega\ \sqrt{\frac{\hbar}{4\pi\diez c\omega \np(\omega)\area}}\
    \op_>(\omega) \ee^{\ii k(\omega)z}.
\end{align}
\end{subequations}
\end{widetext}
Then, Eq.~\eqref{eq:oA_op} is obtained, 
and it has exactly the same form as Eq.~\eqref{eq:A_forward_a}
in the loss-less limit.

\section{Quantum fluctuation in homogeneous dielectric system} \label{app:fluctuation}
In an homogeneous medium with dielectric function $\die(z,\omega) = \die(\omega)$,
from the wave equation \eqref{eq:Maxwell_wave_A},
the vector potential is expressed as
\begin{equation}
\oA^+(z,\omega) = \oA_{>}^+(z,\omega) + \oA_{<}^+(z,\omega),
\end{equation}
where $\oA_>^+$ and $\oA_<^+$ are forward and backward fields defined as
\begin{subequations} \label{eq:A_forward_J} 
\begin{align}
\oA_{>}^+(z,\omega) & = - \muz \int_{-\infty}^z\dd z'\ \frac{\ee^{\ii k(\omega)(z-z')}}{\ii2k(\omega)} \oJN(z',\omega), \\
\oA_{<}^+(z,\omega) & = - \muz \int_z^{\infty}\dd z'\ \frac{\ee^{-\ii k(\omega)(z-z')}}{\ii2k(\omega)} \oJN(z',\omega).
\end{align}
\end{subequations}
Since the noise current density $\oJN(z,\omega)$ has the local correlation
(commutable for different positions $z \neq z'$),
the forward field $\oA_>(z)$ and backward one $\oA_<(z')$
are commutable for $z < z'$
\begin{equation}
[ \oA_>^{+}(z,\omega), \oA_<^-(z',\omega') ] = 0 \quad \text{for $z < z'$}.
\end{equation}
The quantum fluctuation of the vector potential
in the frequency domain is obtained as
\begin{align}
& [ \oA_>^{+}(z,\omega), \oA_>^-(z',\omega') ]
\nonumber \\ &
= \delta(\omega-\omega') \frac{\hbar}{4\pi\diez c\omega\area}
  \frac{\Re[n(\omega)]}{|n(\omega)|^2}
  \ee^{\ii\Re[k(\omega)](z-z') - \Im[k(\omega)]|z-z'|}.
\end{align}
When we define the spatial Fourier transform for wavenumber $q > 0$ as
\begin{equation}
\oA^+_q = \frac{1}{\sqrt{\len}} \int\dd z\ \ee^{-\ii q z} \oA^+_>(z,\omega),
\end{equation}
the quantum fluctuation of this mode in the frequency domain is obtained as
\begin{widetext}
\begin{subequations}
\begin{align}
[ \oA^{+}_q(\omega), \oA^-_{q'}(\omega') ]
& = \delta(\omega-\omega') \delta_{q,q'}
  \frac{2\Im[k(\omega)]}{|q-k(\omega)|^2}
  \frac{\hbar}{4\pi\diez c\omega\area}
  \frac{\Re[n(\omega)]}{|n(\omega)|^2} \\
& = \delta(\omega-\omega') \delta_{q,q'}
  \frac{\hbar}{4\pi\diez c\omega\area}
  \frac{\Re[n(\omega)]}{|n(\omega)|^2}
  \left[
    \int_{-\infty}^0\dd z\ \ee^{\ii[q-k(\omega)]z}
  + \int_0^{\infty}\dd z\ \ee^{\ii[q-k(\omega)^*]z}
  \right].
\end{align}
\end{subequations}
\end{widetext}
In the loss-less limit ($\Im[\die(\omega)] \rightarrow 0$),
we get
\begin{equation}
[ \oA^{+}_q(\omega), \oA^-_{q'}(\omega') ]
= \delta(\omega-\omega') \delta_{q,q'} \delta(q-k(\omega))
  \frac{\hbar}{2\diez c\omega\area}
  \frac{1}{n(\omega)}.
\end{equation}
Then, the equal-time quantum fluctuation of this mode is finally written as
\begin{subequations}
\begin{align}
[ \oA^{+}_{q}, \oA^-_{q'} ]
& = \delta_{q,q'} \int_0^{\infty}\dd\omega\ \delta(q-k(\omega))
    \frac{\hbar}{2\diez c\omega\area}
    \frac{1}{n(\omega)} \nonumber \\
& = \delta_{q,q'}
    \frac{\hbar}{2\diez cq\area}
    \frac{1}{n(\varOmega_q)},
\end{align}
where $\varOmega_q = c q / n(\varOmega_q)$.
Therefore, in the loss-less dielectric media,
the quantum fluctuation of the vector potential
is modified by the factor of $n(\varOmega_q)^{-1}$.
The fluctuations of the other electromagnetic fields are obtained as
\begin{align}
[ \oE^{+}_{q}, \oE^-_{q'} ]
& = \delta_{q,q'} \frac{\hbar cq}{2\diez\area} \frac{1}{n(\varOmega_q)^3}, \\
[ \oB^{+}_q, \oB^-_{q'} ]
& = \delta_{q,q'} \frac{\hbar q}{2\diez c\area} \frac{1}{n(\varOmega_q)}, \\
[ \oD^{+}_{q}, \oD^-_{q'} ]
& = \delta_{q,q'} \frac{\hbar cq}{2\diez\area} n(\varOmega_q).
\end{align}
\end{subequations}
The dependence of these fields on $n(\varOmega_q)$ is exactly the same
as the one discussed in Ref.~\cite{Glauber1991PRA}.
Then, the electromagnetic fields are sub-fluctuant or super-fluctuant
in dielectrics compared to the case in vacuum.

\begin{widetext}
The forward and backward fields \eqref{eq:A_forward_J} are rewritten as
\begin{subequations} \label{eq:A_forward_a} 
\begin{align}
\oA_{>}^+(z,\omega) &
= \sqrt{\frac{\hbar}{4\pi\diez c\omega \Re[n(\omega)]\area}} \frac{\Re[n(\omega)]}{n(\omega)} \ee^{\ii\Re[k(\omega)] z} \oa_>(z,\omega), \\
\oA_{<}^+(z,\omega) &
= \sqrt{\frac{\hbar}{4\pi\diez c\omega \Re[n(\omega)]\area}} \frac{\Re[n(\omega)]}{n(\omega)} \ee^{-\ii\Re[k(\omega)] z} \oa_<(z,\omega),
\end{align}
\end{subequations}
where operators $\oa_{>}$ and $\oa_{<}$ are defined as
\begin{subequations}
\begin{align}
\oa_{>}(z,\omega) &
= \ii \sqrt{\frac{\pi\muz c\area}{\hbar\omega\Re[n(\omega)]}}\ee^{-\Im[k(\omega)]z}
  \int_{-\infty}^{z}\dd z'\ \ee^{-\ii k(\omega) z'} \oJN(z',\omega), \\
\oa_{<}(z,\omega) &
= \ii \sqrt{\frac{\pi\muz c\area}{\hbar\omega\Re[n(\omega)]}}\ee^{\Im[k(\omega)]z}
  \int_z^{\infty}\dd z'\ \ee^{\ii k(\omega) z'} \oJN(z',\omega).
\end{align}
\end{subequations}
\end{widetext}
They correspond to the annihilation operator of a photon in the dielectric medium,
and the commutator is derived as
\begin{align}
& [ \oa_>(z,\omega), \oad_>(z',\omega) ]
\nonumber \\ &
= [ \oa_<(z,\omega), \oad_<(z',\omega) ]
= \delta(\omega-\omega') \ee^{-\Im[k(\omega)]|z-z'|}
\end{align}
In the loss-less limit ($\Im[\die(\omega)] \rightarrow 0$),
they become position-independent
and simply considered as the annihilation operator.

\section{Solution in cavity system by Green's function approach} \label{app:Green}
Let's derive the Green's function $G(z,z',\omega)$ satisfying
\begin{equation}
- \left[ \frac{\partial^2}{\partial z^2} + \frac{\omega^2}{c^2}\die(z,\omega) \right]
G(z,z',\omega) = \delta(z-z').
\end{equation}
where the dielectric function is expressed as Eq.~\eqref{eq:die}.
Obeying the recipe in Ref.~\cite{chew95},
the Green's function $G_{ij}(z,z',\omega)$
($z$ in region $i$ and $z'$ in region $j$) is obtained as
\begin{widetext}
\begin{subequations}
\begin{align}
G_{11}(z,z',\omega)
& = - \frac{1}{\ii2(\omega/c)} \left\{
      \ee^{\ii(\omega/c)|z-z'|}
    + \ee^{-\ii(\omega/c)z} \gR_{21}(\omega) \ee^{-\ii(\omega/c)z'}
    \right\}, \\
G_{21}(z,z',\omega)
& = - \frac{1}{\ii2(\omega/c)}
     \sin[\kp(\omega)(\cavlen-z)] \gT_{21}(\omega) \ee^{-\ii(\omega/c)z'},
     \\
G_{12}(z,z',\omega)
& = - \frac{1}{\ii2\kp(\omega)}\ee^{-\ii(\omega/c)z} \gT_{12}(\omega) \sin[\kp(\omega)(\cavlen-z')],\\
G_{22}(z,z',\omega)
& = - \frac{1}{\ii2\kp(\omega)} \left\{
    \ee^{\ii\kp(\omega)|z-z'|} - \ee^{-\ii\kp(\omega)(z-\cavlen)}\ee^{-\ii\kp(\omega)(z'-\cavlen)}
\right. \nonumber \\ & \quad \left.
    + \frac{\ii2\ee^{\ii\kp(\omega)\cavlen}[1-\ii\varLambda(\omega)-\np(\omega)]}
           {[1-\ii\varLambda(\omega)]\sin[\kp(\omega)\cavlen]+\ii \np(\omega) \cos[\kp(\omega)\cavlen]}
      \sin[\kp(\omega)(\cavlen-z)]\sin[\kp(\omega)(\cavlen-z')]
    \right\} \\
& = - \frac{1}{\ii2\kp(\omega)} \left\{
      \ee^{\ii\kp(\omega)|z-z'|}
    - \ee^{\ii\kp(\omega)z}
      \frac{[1-\ii\varLambda(\omega)-\np(\omega)]\sin[\kp(\omega)(\cavlen-z')]}
           {[1-\ii\varLambda(\omega)]\sin[\kp(\omega)\cavlen]+\ii \np(\omega) \cos[\kp(\omega)\cavlen]}
\right. \nonumber \\ & \quad \left.
    - \ee^{-\ii\kp(z-\cavlen)}
      \frac{[1-\ii\varLambda(\omega)]\sin[\kp(\omega)z']+\ii \np(\omega) \cos[\kp(\omega)z']}
           {[1-\ii\varLambda(\omega)]\sin[\kp(\omega)\cavlen]+\ii \np(\omega) \cos[\kp(\omega)\cavlen]}
    \right\},
\end{align}
\end{subequations}
\begin{subequations}
\begin{align}
\gR_{21}(\omega)
& = \frac{[1+\ii\varLambda(\omega)]\sin[\kp(\omega)\cavlen]-\ii \np(\omega) \cos[\kp(\omega)\cavlen]}
         {[1-\ii\varLambda(\omega)]\sin[\kp(\omega)\cavlen]+\ii \np(\omega) \cos[\kp(\omega)\cavlen]}, \\
\gT_{21}(\omega)
& = \frac{2}{[1-\ii\varLambda(\omega)]\sin[\kp(\omega)\cavlen]+\ii \np(\omega) \cos[\kp(\omega)\cavlen]}, \\
\gT_{12}(\omega)
& = \frac{2\np(\omega)}
         {[1-\ii\varLambda(\omega)]\sin[\kp(\omega)\cavlen]+\ii \np(\omega) \cos[\kp(\omega)\cavlen]}.
\end{align}
\end{subequations}
\end{widetext}
The derivation is as follows.
When a source exists in region 1, the Green's function
can be supposed as follows:
\begin{equation} \label{eq:Green} 
G_{j1}(z,z',\omega) = - \frac{\GG_{j1}(z,z',\omega)}{\ii2(\omega/c)},
\end{equation}
\begin{subequations}
\begin{align}
\GG_{11}(z,z',\omega)
& = \ee^{\ii(\omega/c)|z-z'|} + \ee^{-\ii(\omega/c)z} B_{11}(z'), \\
\GG_{21}(z,z',\omega)
& = \ee^{\ii\kp(\omega)(z-\cavlen)} F_{21}(z') + \ee^{-\ii\kp(\omega)(z-\cavlen)} B_{21}(z').
\end{align}
\end{subequations}
We get a boundary condition at $z = \cavlen$:
\begin{subequations}
\begin{equation}
F_{21}(z') + B_{21}(z') = 0.
\end{equation}
At $z = 0$, we also get
\begin{equation}
\ee^{-\ii(\omega/c)z'} + B_{11}(z')
= \ee^{-\ii\kp(\omega)\cavlen} F_{21}(z') + \ee^{\ii\kp(\omega)\cavlen} B_{21}(z'),
\end{equation}
\begin{align}
& \ee^{-\ii(\omega/c)z'} - B_{11}(z')
\nonumber \\ &
- \np(\omega) \left[ \ee^{-\ii\kp(\omega)\cavlen} F_{21}(z') - \ee^{\ii\kp(\omega)\cavlen} B_{21}(z') \right]
\nonumber \\ & \quad
= - \ii\varLambda(\omega) \left[ \ee^{-\ii(\omega/c)z'} + B_{11}(z') \right].
\end{align}
\end{subequations}
The third condition is obtained by the boundary condition \eqref{eq:MBC_B_cavity}
or integrating Eq.~\eqref{eq:Green}.
Solving them, we get $G_{11}$ and $G_{21}$.

When a source exists at region 2, we can suppose
\begin{equation}
G_{j2}(z,z',\omega) = - \frac{\GG_{j2}(z,z',\omega)}{\ii2\kp(\omega)},
\end{equation}
\begin{subequations}
\begin{align}
\GG_{12}(z,z',\omega)
& = \ee^{-\ii(\omega/c)z} B_{12}(z'), \\
\GG_{22}(z,z',\omega)
& = \ee^{\ii\kp(\omega)|z-z'|}
\nonumber \\ & \quad
  + \ee^{\ii\kp(\omega)z} F_{22}(z') + \ee^{-\ii\kp(\omega)(z-\cavlen)} B_{22}(z').
\end{align}
\end{subequations}
At $z = \cavlen$, we get
\begin{subequations}
\begin{equation}
\ee^{\ii\kp(\omega)(\cavlen-z')} + \ee^{\ii\kp(\omega)\cavlen} F_{22}(z') + B_{22}(z') = 0.
\end{equation}
Further, at $z = 0$
\begin{equation}
\ee^{\ii\kp(\omega)z'} + F_{22}(z') + \ee^{\ii\kp(\omega)\cavlen} B_{22}(z') = B_{12}(z'),
\end{equation}
\begin{align}&
B_{12}(z') + \np(\omega) \left[ - \ee^{\ii\kp(\omega)z'} + F_{22}(z') - \ee^{\ii\kp\cavlen} B_{22}(z') \right]
\nonumber \\ & \quad
= \ii\varLambda(\omega) B_{12}(z').
\end{align}
\end{subequations}
Then, $G_{22}$ and $G_{12}$ are obtained.

By using the Green's function, the vector potential is obtained in region 1 as
\begin{widetext}
\begin{multline} \label{eq:A1_Green} 
\oA^+_1(z,\omega)
= \ee^{\ii(\omega/c)z} \oA^+_{1>}(z,\omega)
+ \ee^{-\ii(\omega/c)z} \gR_{21} \oA_{1>}^+(0,\omega) \\
- \frac{\muz}{\ii2(\omega/c)}\int_z^0\dd z'\ \ee^{-\ii(\omega/c)(z-z')} \oJN(z')
+ \muz\int_0^{\cavlen}\dd z'\ G_{12}(z,z',\omega) \oJN(z').
\end{multline}
\end{widetext}
where we define the incoming field as
\begin{equation}
\oA_{1>}^+(z,\omega) = - \frac{\muz}{\ii2(\omega/c)} \int_{-\infty}^z\dd z'\ \ee^{-\ii(\omega/c)z'} \oJN(z').
\end{equation}
Here, as discussed in Ref.~\cite{knoll01},
we can focus only on the incoming and outgoing fields at the boundary $z = 0$,
because the contribution such as the third term in Eq.~\eqref{eq:A1_Green}
can be neglected in the loss-less limit
(if there is no dissipation, we need not consider the noise operator
for the propagation in vacuum).
Further, the last term can also be neglected in the loss-less limit,
then the vector potential in region 1 can be expressed as
\begin{equation}
\oA^+_1(z,\omega)
= \ee^{\ii(\omega/c)z} \oA^+_{1>}(\omega)
+ \ee^{-\ii(\omega/c)z} \oA_{1<}^+(\omega).
\end{equation}
The first term corresponds to the incoming field
$\oA^+_{1>}(\omega) = \oA^+_{1>}(0,\omega)$,
and the second term is the outgoing field represented as
\begin{equation}
\oA_{1<}^+(\omega) = \gR_{21} \oA_{1>}^+(0,\omega).
\end{equation}
On the other hand, the vector potential in region 2 is obtained
in the loss-less limit as
\begin{equation}
\oA^+_2(z,\omega)
= \gT_{21}\oA^+_{1>}(\omega) \sin[\kp(\omega)(\cavlen-z)].
\end{equation}
Then, we get
\begin{equation}
\oA^+_2(\omega) = \gT_{21}\oA^+_{1>}(\omega).
\end{equation}
These fields $\oA^+_{1>}(\omega)$, $\oA^+_{1<}(\omega)$, $\oA^+_{2}(\omega)$
certainly satisfy the Maxwell's boundary conditions \eqref{eq:MBC_cavity}.


\end{document}